\documentclass{article}
\usepackage{graphicx,amsmath,epsfig,fancyheadings,datetime}
\usepackage[comma,sort]{natbib}
\usepackage[hang]{subfigure}

\usepackage{pst-all}

\newenvironment{narrow}[2]{%
   \begin{list}{}{%
	\setlength{\topsep}{0pt}%
	\setlength{\leftmargin}{#1}%
	\setlength{\rightmargin}{#2}%
	\setlength{\listparindent}{\parindent}%
	\setlength{\itemindent}{\parindent}%
	\setlength{\parsep}{\parskip}}%
   \item[]}{\end{list}}

\begin{document}           

\renewcommand{\baselinestretch}{2}
\normalsize

\title{The dynamics of Machiavellian intelligence}
\author{Sergey Gavrilets$^{*\dag}$ and Aaron Vose$^{\ddag}$\\
Departments of $^*$Ecology and Evolutionary Biology,
$^{\dag}$Mathematics,\\ and 
$^{\ddag}$Computer Science, University of Tennessee,\
Knoxville, TN 37996, USA.\\}

\maketitle

{\small {\bf Abstract}. The ``Machiavellian intelligence'' hypothesis (also referred to as the
``social brain'' hypothesis) posits that large brains and distinctive cognitive 
abilities of humans have evolved via a spiraling arms race in which social competitors 
developed increasingly sophisticated ``Machiavellian'' strategies. 
Here we  build a mathematical model aiming to explore this hypothesis. In the model, genes control 
brains which invent and learn strategies (memes) which are used by males to gain advantage in 
competition for mates. We show that the dynamics of intelligence has three distinct phases.
During  the dormant phase only newly invented memes are present in the population.
During the cognitive explosion phase the population's meme count and the learning ability, cerebral 
capacity (controlling the number of different memes that the brain can learn socially and use), and 
Machiavellian fitness of individuals rapidly increase in a runaway fashion.
During the saturation phase natural selection resulting from the costs of having large brains checks
further increases in cognitive abilities. Overall, our results suggest that the mechanisms
underlying the ``Machiavellian intelligence'' hypothesis
can indeed result in the evolution of significant cognitive abilities on the time scale 
of 10 to 20 thousand generations.
We show that cerebral capacity evolves faster and to a larger degree than learning ability.
The increase in brain size results in a significant reduction in viability thus creating
conditions that favor rapid evolution of the mechanisms reducing the costs of having
large brains (such as postponing much of the brain growth to after birth and reduction of the guts).
Our model suggests that there may be a tendency
towards a reduction in cognitive abilities (driven by the costs of having a large
brain) as the reproductive advantage of having a large brain decreases
and the exposure to memes increases in modern societies.}\\

There are many features that make us a ``uniquely unique species'' but the most crucial of
them are related to the size and complexity of our brain \citep{str05,gea05,rot05}. 
The brain size in {\em Homo sapiens} 
increased in a runaway fashion over a period of a couple hundred thousand years but then 
stabilized or even slightly declined in the last 35-50 thousand years \citep{ruf97,str05,gea05}. 
In humans, the brain is
very expensive metabolically: it represents about 2\% of the body's weight but utilizes about
20\% of total body metabolism at rest \citep{hol96}. The two burning questions are what factors drove the 
evolution of brain size and why our ancestors 50,000 years ago needed the brains they had.
A number of potential answers have been hotly debated focusing on the effects of climatic \citep{vrb95}, 
ecological \citep{rus04}, and 
social factors. One controversial set of ideas \citep{hum76,byr88,ale90,whi97,dun98,dun03,fli05,str05,gea05,rot05} 
coming under the rubric of the ``Machiavellian intelligence'' or 
``social brain'' hypothesis identifies selective forces resulting from social
competitive interactions as the most important factor in the evolution of hominids, who at some point in
the past became an ecologically dominant species \citep{ale90,fli05}. These forces selected for more and more
effective strategies of achieving social success (including deception, manipulation, alliance 
formation, exploitation of the expertise of others, etc.) and for ability to learn and use them.
The social success translated into reproductive success \citep{bet86,bet93,zer03} selecting
for larger and more complex brains. Once a tool for inventing, learning, and using these 
strategies (i.e., a complex brain) is in place, it can be used for a variety of other purposes 
including coping with environmental, ecological, technological, linguistic, and other challenges. 

Although these ideas are by now well-appreciated by many, and some components of the general scenario
are supported by data \citep{byr04,paw98,saw92,saw97}, 
verbal arguments and generalization 
from limited data alone are not enough to establish their general plausibility and predict the 
relevant time-scales and expected dynamic patterns. Here we attempt to shed some light on these questions using a
stochastic individual-based explicit-genetic model.

\section{Model}

We consider a sexual diploid population, and focus on socially learned strategies (memes) 
used by males to gain advantage in competition for mates. As a first step, 
we neglect analogous processes in females (both for simplicity and because sexual selection
in females is expected to be much less intense than in males). 
Genes control the learning ability $a$ and the cerebral capacity $c$ of the brain 
which in turn control how easily a brain learns new strategies (memes) and how many memes a brain 
can host, respectively. Both $a$ and $c$ are treated as additive quantitative characters.
That is, each trait value is found by summing up the contributions of the corresponding
alleles and then normalizing the result. Learning ability $a$ is normalized to be between 
$0$ and $1$, and cerebral capacity $c$ is normalized to be between $0$ and a positive integer 
$c_{\max}$. The loci controlling the two traits are independent, unlinked, diallelic, and 
have equal effects. Both traits are viewed to be directly related to
brain size and complexity and are assumed to be under direct viability selection towards 
$0$. This selection reflects costs (e.g., energetic or due to increased death at childbirth) 
of having large brains. Note that setting the optimum values at $0$ does not mean that having 
no brain at all is optimum but rather reflects a scale chosen. Individuals surviving to
adulthood experience density-dependent mortality maintaining the population size
close to a carrying capacity $K$.

Memes are invented and forgotten by individuals at small constant rates. 
Each meme is characterized by its Machiavellian fitness $\mu$ and complexity $\pi$
($0 \leq \mu,\pi \leq 1$). 
The former contributes to a male's fitness in between-male 
competitive interactions, while the latter defines how easily the meme can be learned.
The correlation $\rho$ between $\mu$ and $\pi$ in newly invented memes is positive reflecting 
the idea that more advantageous memes are, generally, more complex and more difficult to 
learn. The rate of learning a meme is directly proportional to learning 
ability $a$, inversely proportional to the meme's complexity $\pi$, and declines with the ratio 
$n/c$ where $n$ is the number of memes already learnt by the brain.

The Machiavellian fitness $m$ of a male is given by the sum of Machiavellian fitnesses 
of the memes he has learnt. This implies that fitness increases with the number of memes learnt.
The probability that a contest between two males is won by a specific male is
given by an S-shaped function of the corresponding difference in their Machiavellian fitnesses. 
The male's mating rate increases with the average proportion of contests won.
The strength of sexual selection in males is characterized by a parameter $f_{\max}$ measuring 
the number of females fertilized by a male who wins all contests. 
The importance of competition for mating success among males in the model captures
another unique feature of hominids - that mating is possible at most times and that the 
possibility of continual sexual provocation and competition between males is very high \citep{cha53}.
Offspring are produced with account of recombination, segregation, and mutation.

In our model, there are two types of selection: among memes and among genes. Although interrelated,
they operate at different time-scales - fast for memes and slow for genes.
To adequately capture this important feature of gene-culture coevolution, 
we use an event-driven modeling framework
in which time is treated as continuous (see {\bf Methods}). Our simulations start with a population of individuals having 
zero learning ability and cerebral capacity.
The population size varied between 50 and 150 individuals which is compatible with social group sizes in 
hominids \citep{dun03}.

\section{Numerical results and biological interpretations}

\begin{figure}[t]
\centering

\subfigure[]{ 	\includegraphics[width=2in]{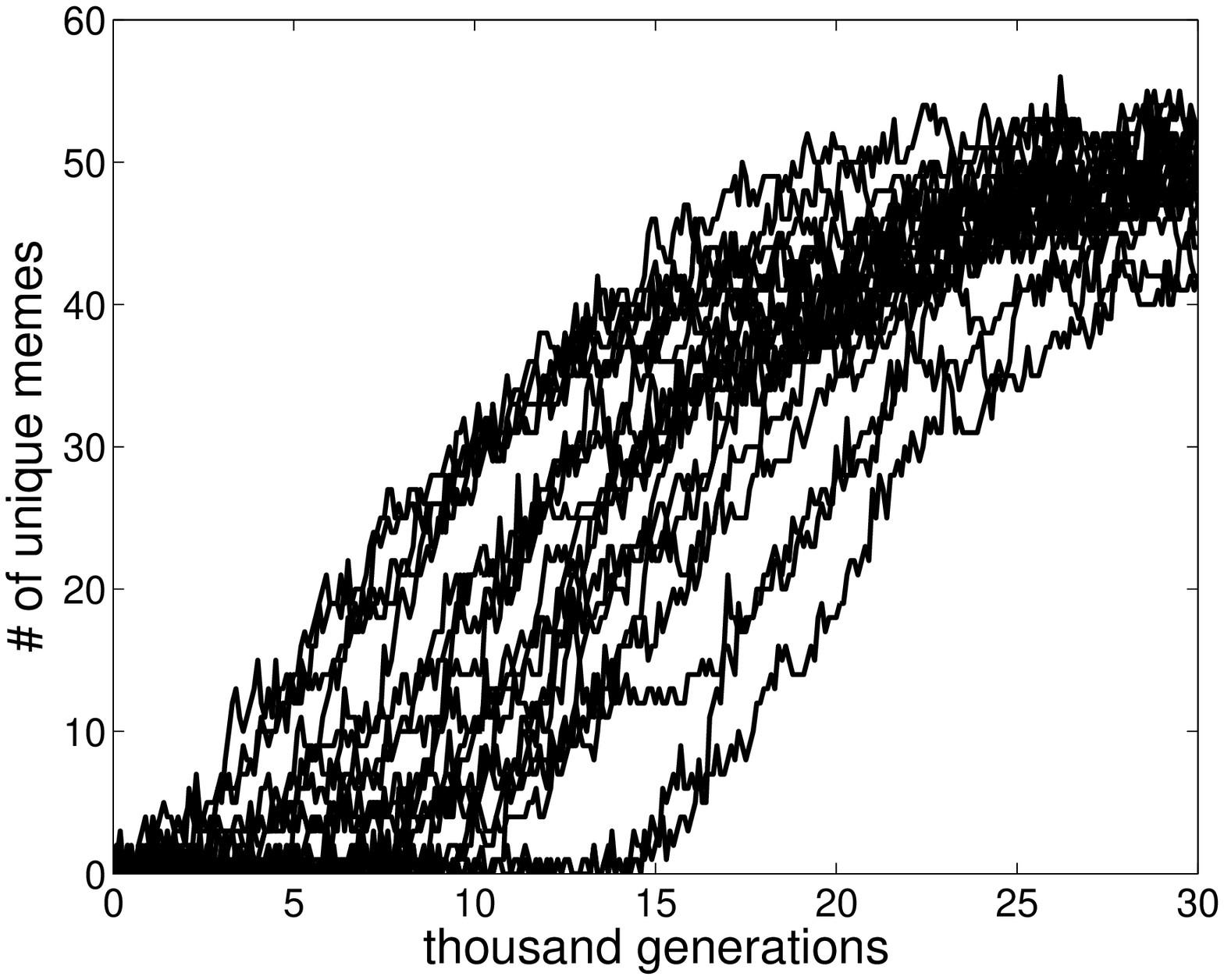}}%
 		\hspace{.5cm}%
\subfigure[]{	\includegraphics[width=2in]{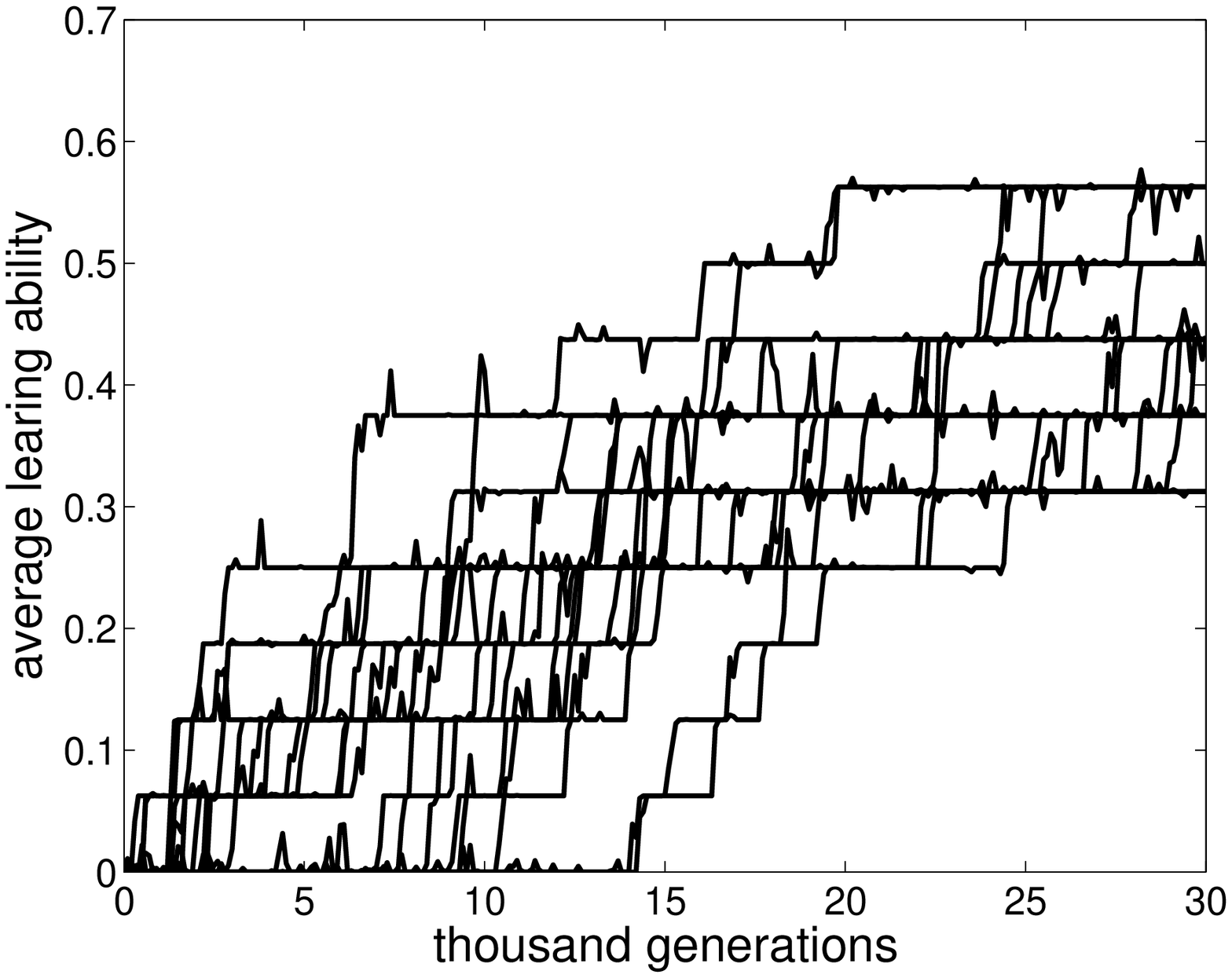}}\\
\subfigure[]{ 	\includegraphics[width=2in]{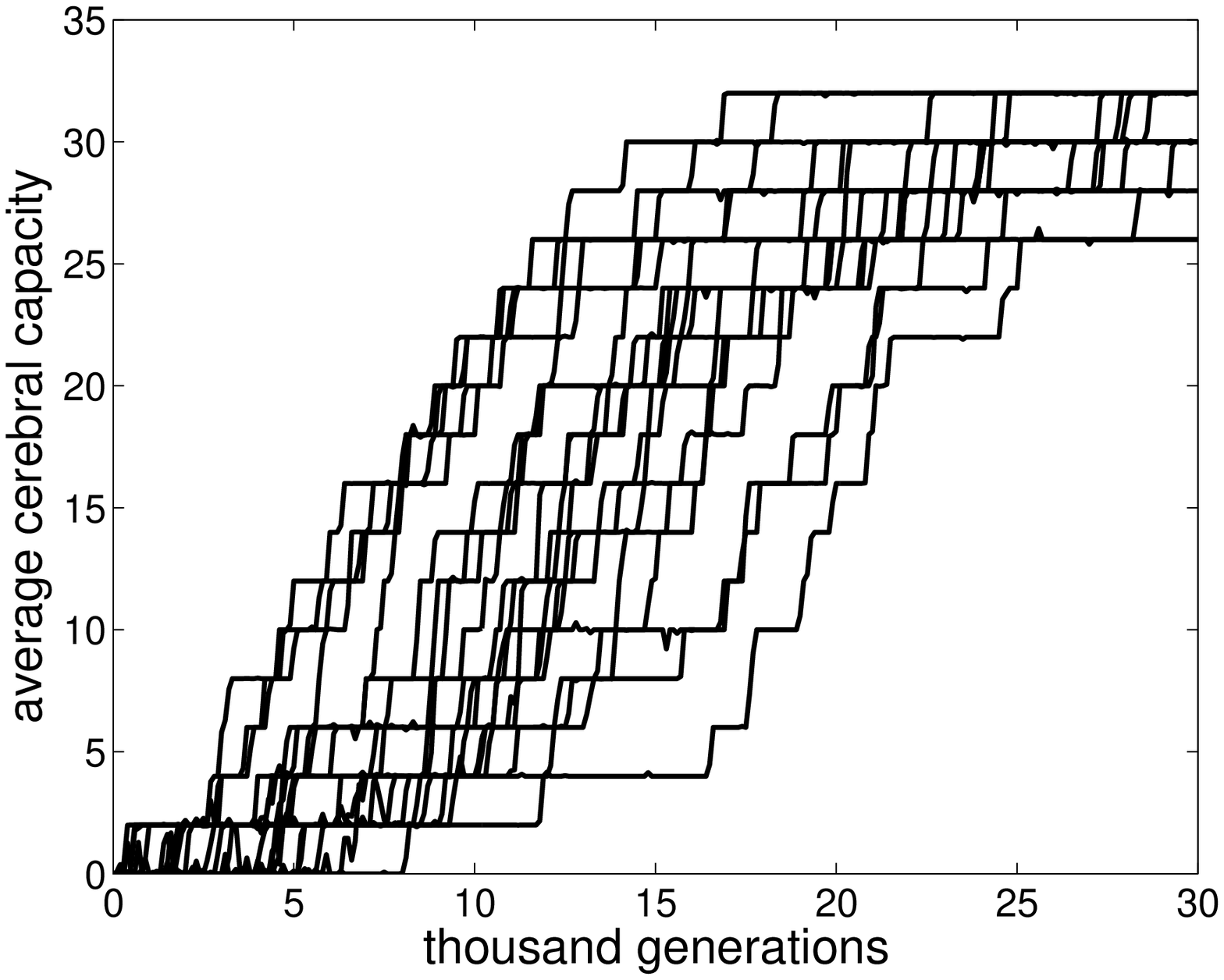}}%
		\hspace{.5cm}%
\subfigure[]{	\includegraphics[width=2in]{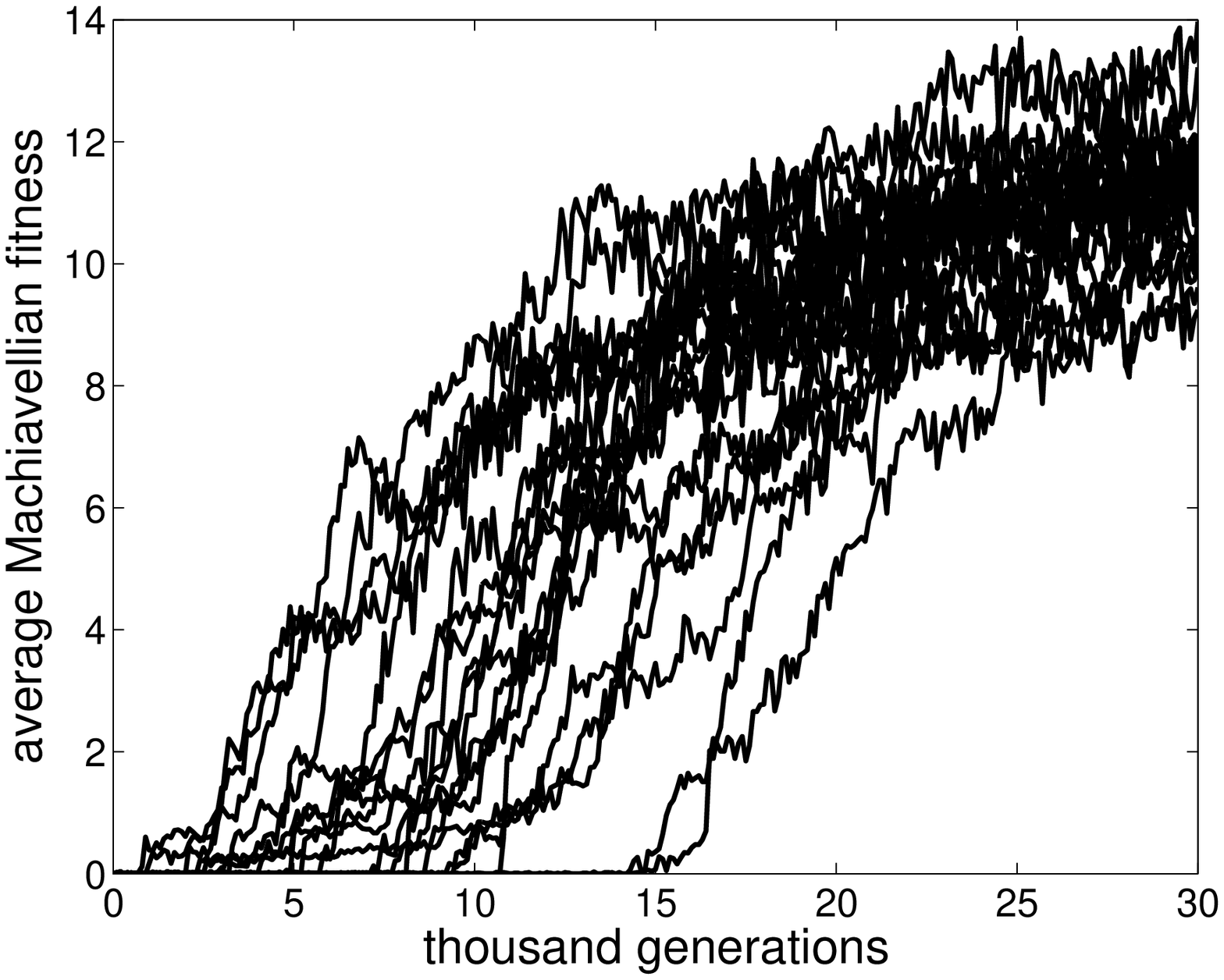}}
\caption{The dynamics of (a)~the number of unique memes, (b)~the average learning ability, 
(c)~the average cerebral capacity, and (d)~the average Machiavellian fitness in 20 runs with a default
set of parameter values ($L=16, K=100, c_{\max}=32, f_{\max}=10, \rho=0.5$).
}
\label{20runs}
\end{figure}

Figure~\ref{20runs} shows the dynamics of the number of unique memes, the average learning ability,
the average cerebral capacity, and the average Machiavellian fitness of males in 20 runs with a default
set of parameter values. Each of these characteristics stays close to zero for several thousand
generations during the ``dormant phase'' and then suddenly starts rapidly increasing in a process 
that we will refer to as ``cognitive explosion.'' Cognitive explosion ends when natural selection
stops further increase in cognitive abilities due to the costs of having large brains, and the system
enters the ``saturation phase.'' During the whole process the population stays genetically monomorphic
except during relatively short periods of ``selective sweeps'' when mutant alleles go to fixation
(data not shown).

The dynamics before and at the onset of the cognitive explosion can be understood as follows. 
Nonzero learning ability $a$ and cerebral capacity $c$ are advantageous only if the individual has 
both of them simultaneously and, during his life time, learns a meme (or memes) from other 
individuals. Otherwise, individuals with $a>0$ and/or $c>0$ have reduced fitness due to decreased viability.
The resulting fitness landscape resembles that in models of compensatory mutations \citep{kim85,gav04} 
where a deleterious
mutation in one locus can be compensated later on by an advantagenous mutation in a different locus.
During the dormant phase, one of the traits (i.e., learning ability or cerebral capacity) can sporadically
deviate away from zero by mutation and random genetic drift in spite of this deviation causing a reduction 
in fitness. Cognitive explosion takes place when individuals with nonzero values of one trait 
are maintained in the population sufficiently long for mutations changing the value
of the other trait in their offspring to occur and when both sets of new genes are maintained 
in the population long enough for the individuals to learn new memes and start enjoying a fitness advantage. 

\begin{figure}[t]
\centering
\includegraphics[width=5in]{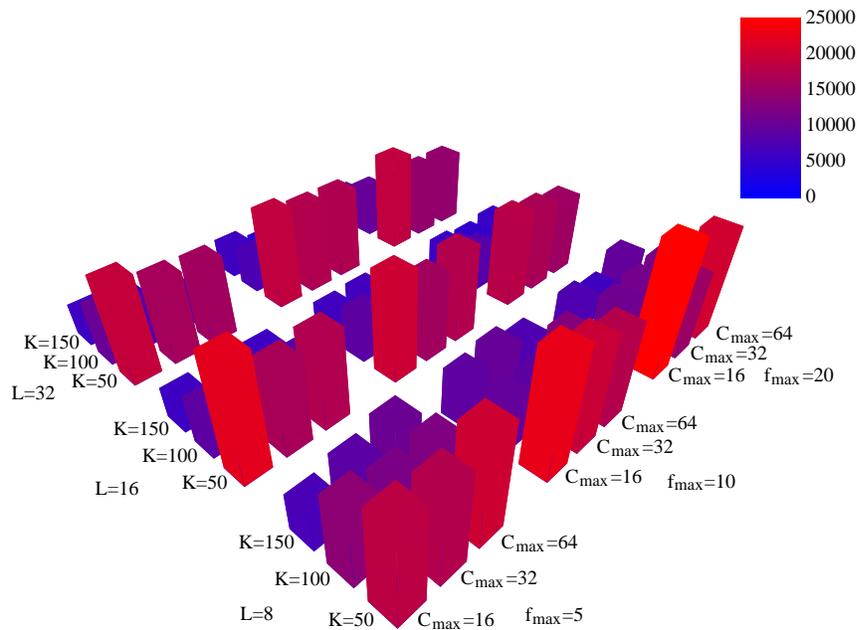}
\caption{The median time to cognitive explosion with $\rho=0.5$. 
}
\label{time2explosion}
\end{figure}

The onset of cognitive explosion depends on parameter values and varies from run to run.
Figure~\ref{time2explosion} illustrates the dependence of the median time $T$ to
cognitive explosion on the population carrying capacity $K$, the number of loci $L$ underlying each trait, 
the maximum cerebral capacity $c_{\max}$, and the maximum mating group size $f_{\max}$ when the
correlation $\rho$ between meme fitness and complexity is 0.5.
This Figure as well as a statistical analysis based on the Cox proportional-hazard regression \citep{and82} 
(data not shown) show that $T$ decreases with increasing $K,L,c_{\max},f_{\max}$ and decreasing $\rho$.
The effects of $K$ and $L$ are the most pronounced which is compatible with the idea that the process of fixation
of compensatory mutations is mostly limited by the availability of new genetic variation \citep{kim85,gav04}. 
With realistic parameter values the waiting time until the onset of
cognitive explosion is on the order of 5-25 thousand generations.

\begin{figure}
\centering
\includegraphics[width=2in]{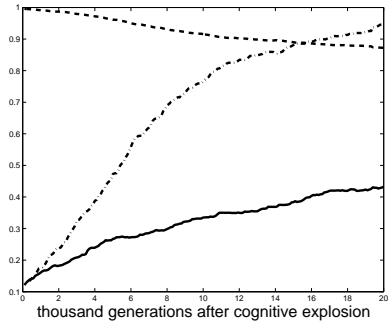}
\caption{The dynamics of the learning ability $a$ (solid line), normalized
cerebral capacity $c/c_{\max}$  (dashed-dot line) and  viability $v$ (dashed line) averaged
over 20 runs shown in Figure 1.
}
\label{average20}
\end{figure}

After the onset of the cognitive explosion, further increases in $a$ and $c$ by mutation are advantageous
because they allow individuals to learn more memes and, thus, achieve higher Machiavellian fitness and
mating rate. 
A novel feature of our framework is an explicit consideration of the dynamics of cerebral capacity $c$
which we defined as a measure of the number of memes that the brain can host. In previous models,
$c$ was implicitly assumed to be fixed at one \citep{cav81,boy85} or infinity \citep{hig00}.
During the cognitive explosion phase, cerebral capacity $c$ typically evolves faster and achieves 
higher values than learning ability $a$ (see Figure~\ref{average20}).
This observation suggests that higher values of cerebral capacity are more important 
than high learning ability and that there is more potential
for improving the latter than the former.
Evolution of cognitive abilities results in a significant reduction in individual viability 
(Figure~\ref{average20}) thus creating conditions for the evolution of mechanisms that would
reduce costs of having large brains such as postponing much of the brain growth to after birth \citep{str05} 
and reduction of the guts \citep{aie95}.
In our model, more complex memes provide more fitness benefit to individuals. However,
the complexity of memes present in the population does not increase but, on the contrary, decreases 
in time (data not shown). This happens as a result of intense competition among memes: while complex 
memes give advantage to individuals on a longer (biological) time-scale, they lose competition to 
simpler memes on a shorter (social) time-scale.

\begin{figure}
\begin{narrow}{-.6in}{-.6in}
\centering

\subfigure[]{ 	\includegraphics[width=3.5in]{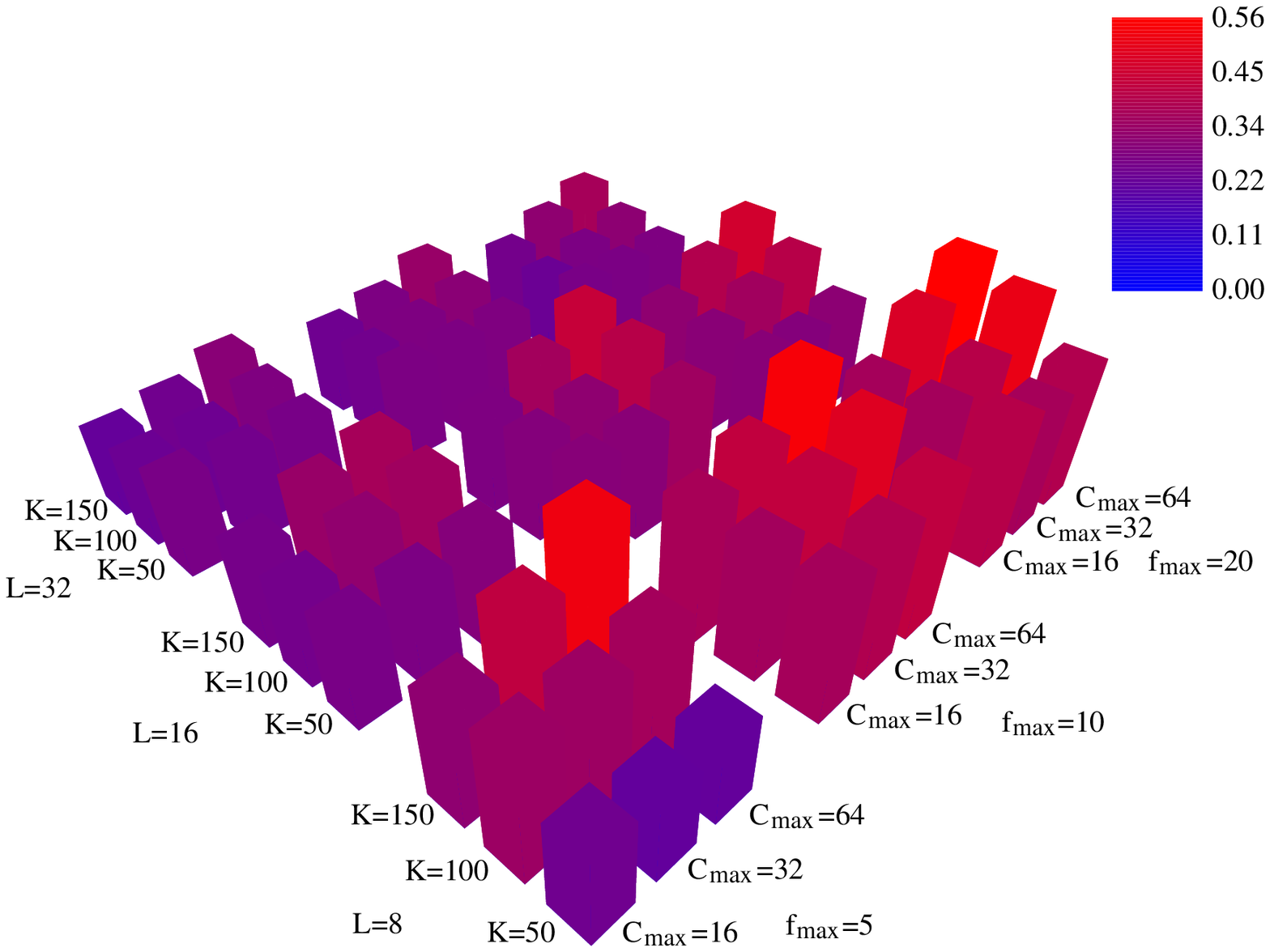}}%
 		\hspace{.1cm}%
\subfigure[]{	\includegraphics[width=3.5in]{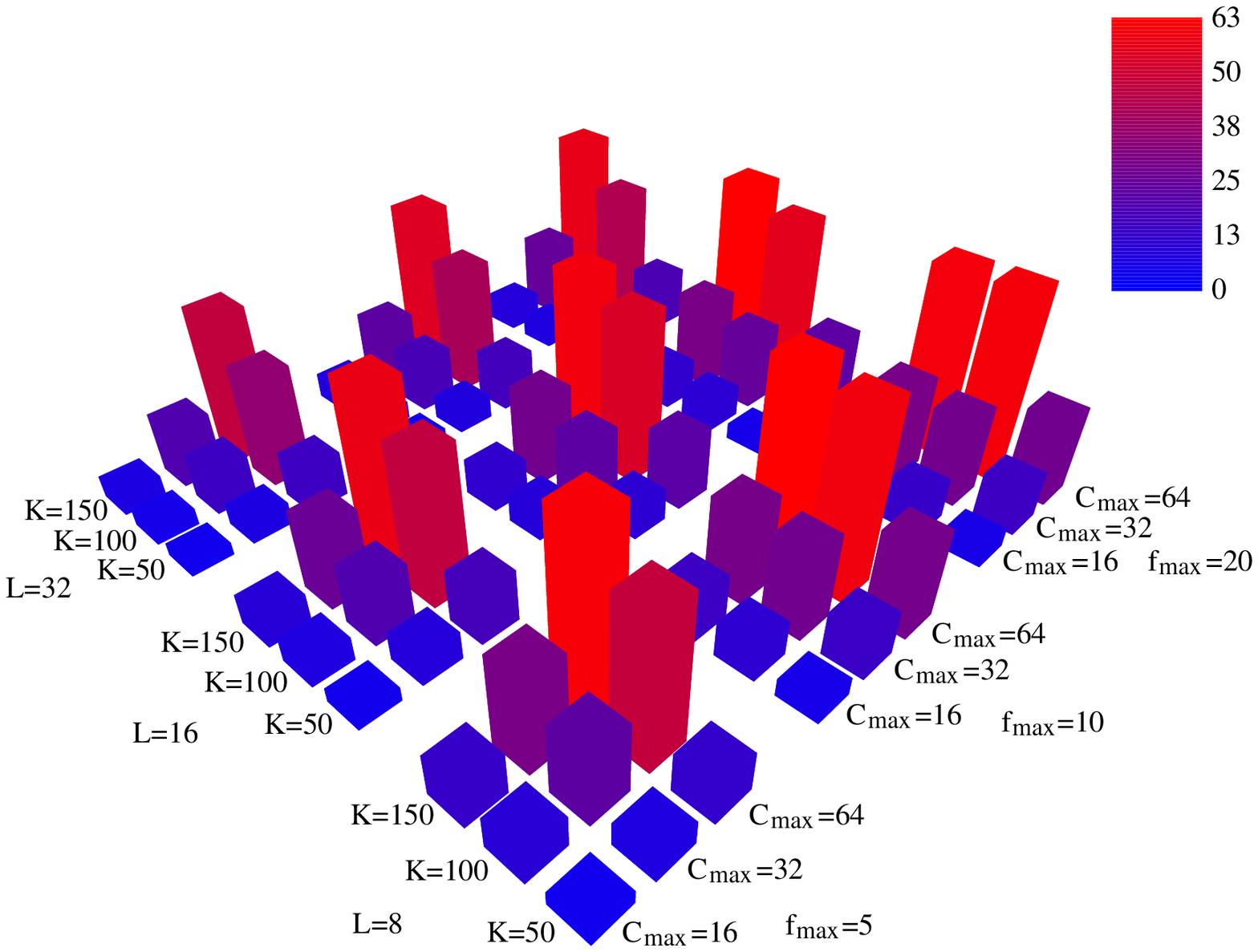}}\\
\subfigure[]{ 	\includegraphics[width=3.5in]{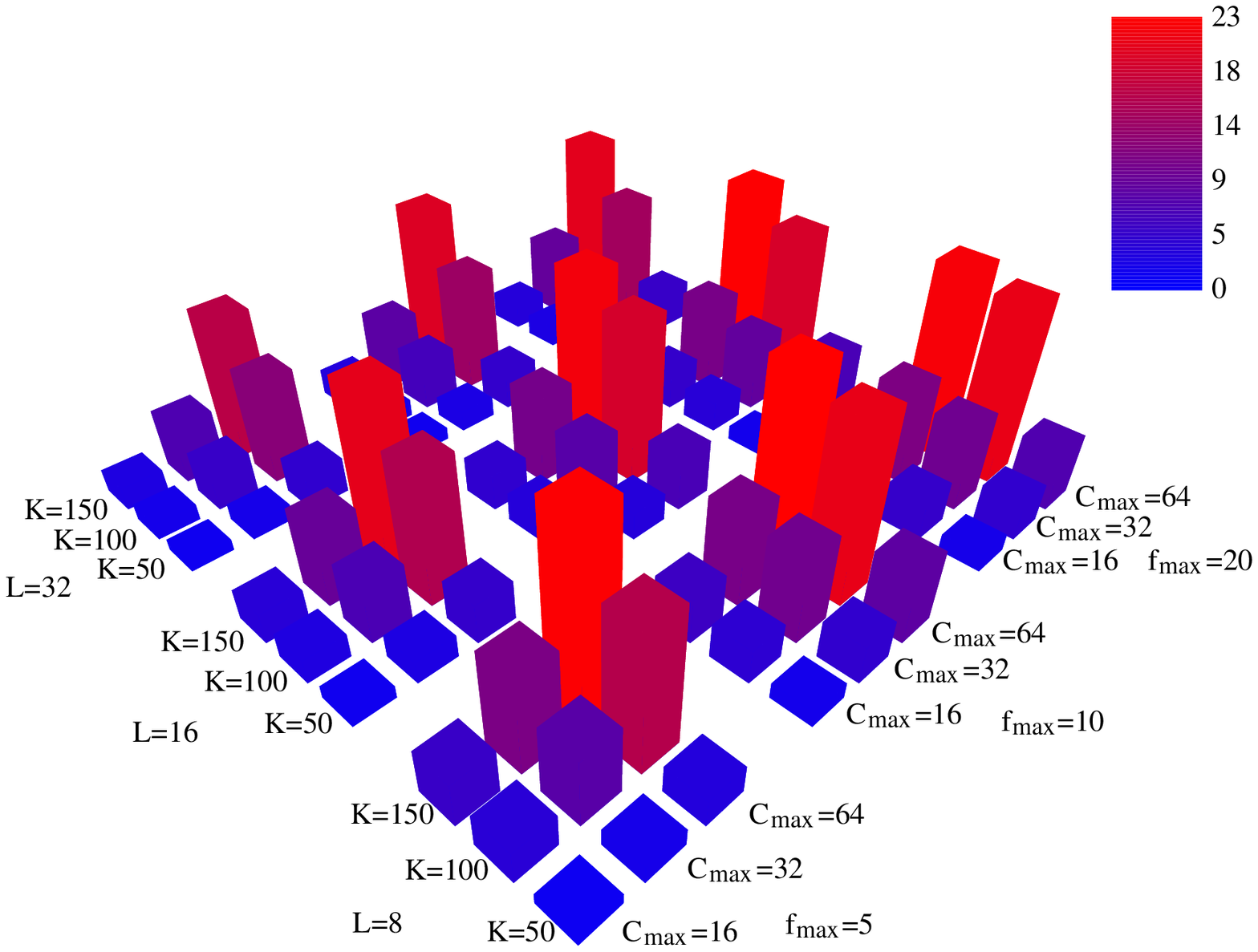}}%
		\hspace{.1cm}%
\subfigure[]{	\includegraphics[width=3.5in]{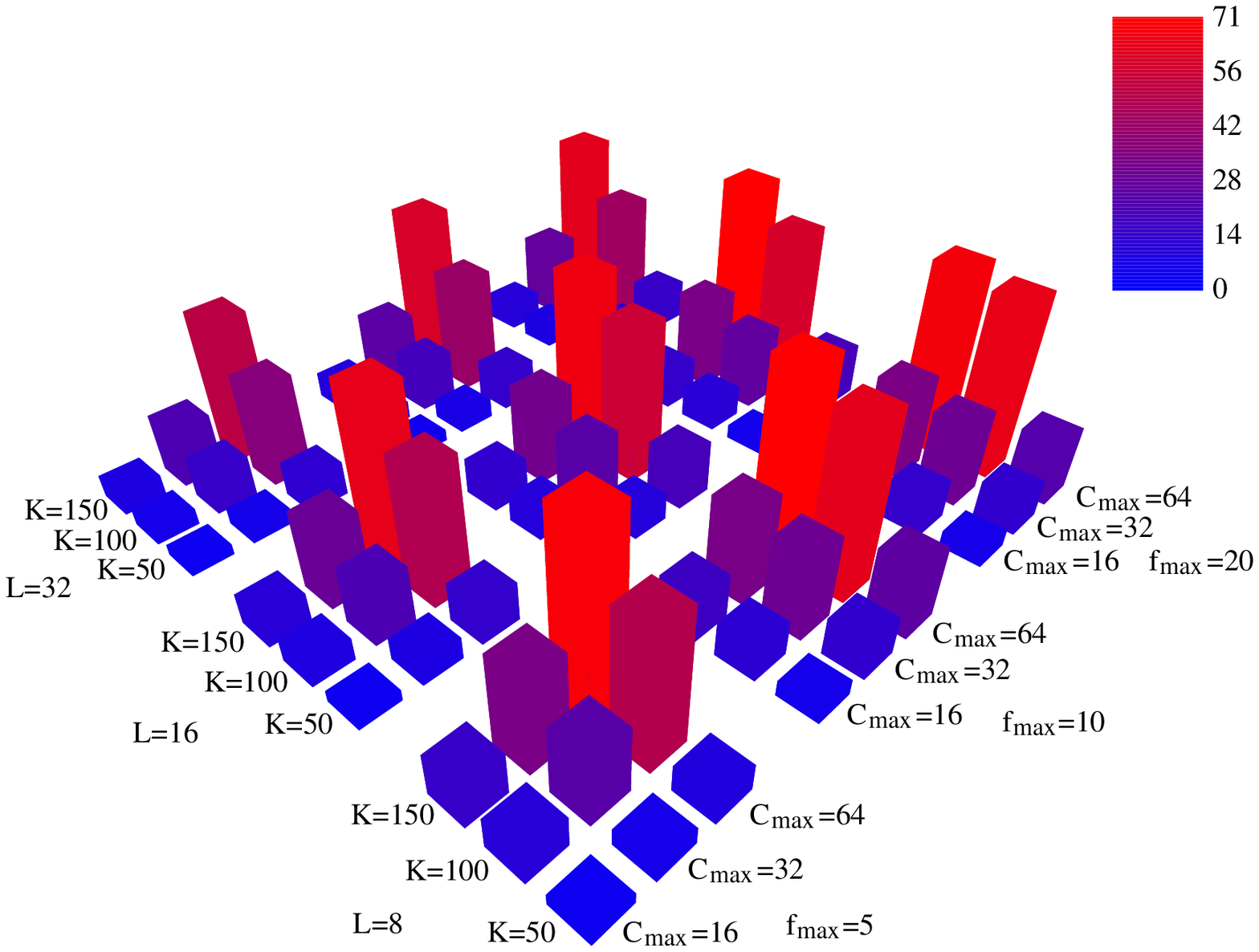}}
\end{narrow}
\caption{The characteristics of the population 8,000 generations after the cognitive explosion:
(a)~the average learning ability,
(b)~the average cerebral capacity, 
(c)~the average Machiavellian fitness, and 
(d)~the average number of memes per individual. $\rho=0.5$.
}
\label{8K}
\end{figure}

In our simulations, the cognitive explosion phase lasts until the cerebral capacity reaches 
a maximum possible level while the learning ability appears to equilibrate at an intermediate 
level determined by a balance of reduced viability and increased mating success of individuals 
having big brains. Figure~\ref{8K} illustrates the state of the population reached in 8,000
generations after the cognitive explosion. This Figure and an analysis of variance (data not shown)
shows that 
the average learning ability, cerebral capacity, Machiavellian fitness, and the number of memes 
per individual all increase with $c_{\max},f_{\max}$, and $K$ and decrease with $\rho$ and $L$. 
The negative effect of the number of loci $L$ on the characteristics of cognitive abilities is 
explained by the fact that more loci means weaker selection on each individual locus and, thus, 
weaker evolutionary response. 
Both the average Machiavellian fitness (Figure~\ref{8K}-c) and the average number of memes per 
individual (Figure~\ref{8K}-d) correlate almost perfectly with the average cerebral capacity 
(Figure~\ref{8K}-b) with the corresponding coefficients of correlation being $0.995$.
Overall, the simulations show that significant values of $c$ and $a$
can be achieved within 5 to 10 thousand generations after the onset of cognitive explosion.

It has been argued that throughout most of human history, success in social competition translated into reproductive 
success with the most powerful men enjoying a disproportionate share of women and offspring \citep{bet86,bet93,zer03}.
In our model, this effect is characterized by parameter $f_{\max}$ measuring the mating 
group size of a male who wins all between-male contests. This parameter strongly affects the levels of learning ability 
and cerebral capacity achieved in the population. For example, with $L=32, c_{\max}=64, K=150$ and
$\rho=0.5$ as $f_{\max}$ decreases from 20 to 10 to 5, 
the average learning ability decreases from 0.36  to 0.33  to 0.30 (Figure~\ref{8K}-a) while the
average cerebral capacity decreases from 57.0 to 54.5 to 47.5 (Figure~\ref{8K}-b).
This suggests that as the extent to which social success translates into reproductive success 
declines in modern societies, cognitive abilities are expected to be significantly reduced by natural 
selection. We also expect that as the number of memes in the population dramatically increases,
learning ability will decrease further (because given a sufficiently large exposure to memes, 
they will be learnt even by individuals with relatively low learning abilities).

%

\section{Discussion}

Here, we have built an explicit-genetic, individual-based, stochastic mathematical model of 
the coevolution of genes and memes aiming to explore the hypothesis of ``Machiavellian intelligence.''
In the model, genes control the learning ability and cerebral capacity of brains which 
invent and learn strategies (memes) which are used by males to gain advantage in competition for mates. 
Overall, our results suggest that the mechanisms underlying this hypothesis
can indeed result in a significant increase in the brain size and in the evolution 
of significant cognitive abilities on the time scale  of 10 to 20 thousand generations.

We show that in our model the dynamics of intelligence has three distinct phases.
During  the dormant phase only newly invented memes are present in the population. These memes are not
learned by other individuals.
During the cognitive explosion phase the population's meme count and the learning ability, cerebral 
capacity, and Machiavellian fitness of individuals rapidly increase in a runaway fashion.
During the saturation phase natural selection resulting from the costs of having large brains checks
further increases in cognitive abilities. 
Both the learning ability and cerebral capacity are subject to negative natural selection due to costs 
of having large brains, but having nonzero values of both traits is necessary for learning and using 
different memes. The process of transition from the dormant phase to the cognitive explosion phase is 
somewhat similar to that of the fixation of a compensatory mutation when higher fitness is achieved by 
fixing two mutations each of which is deleterious by itself. As in the case of compensatory mutations, 
the transition from the dormant phase to the cognitive explosion phase is mostly limited by the 
availability of new genetic variation. 
The levels of cognitive abilities achieved during the cognitive explosion phase increase with the intensity
of competition for mates among males and decrease with the number of loci controlling the brain
size. The latter effect is explained by the fact that a larger number of loci implies weaker selection
on each individual locus. 
In our model, evolutionary processes occur at two different time-scales: fast for memes and slow
for genes. More complex memes provide more fitness benefit to individuals. However, during the cognitive 
explosion phase
the complexity of memes present in the population does not increase but, on the contrary, decreases 
in time. This happens as a result of intense competition among memes: while complex 
memes give advantage to individuals on a slow (biological) time-scale, they lose competition to 
simpler memes on a fast (social) time-scale because they are more difficult to learn.
The increase in brain size results in a significant reduction in viability thus creating
conditions that favor rapid evolution of the mechanisms reducing the costs of having
large brains. One such mechanism is postponing much of the brain growth to after birth \citep{str05}
while another is reduction of the guts \citep{aie95}.
Our model 
suggests that there may be a tendency towards 
a reduction in cognitive abilities (driven by the costs of having a large
brain) as the reproductive advantage of having a large brain decreases
and the exposure to memes increases in modern societies.

The model studied here is based on a novel notion of ``cerebral capacity'' as a measure of the number
of different memes/ideas/strategies that the brain can learn socially and use. This measure
is analogous to ``carrying capacity'' used in ecology to characterize the number of individuals
that can survive in a given ecological niche. During the cognitive explosion phase,
the cerebral capacity evolves faster and to a larger degree than learning ability.
Both the average Machiavellian fitness and the average number of memes per 
individual achieved during the cognitive explosion phase are largely controlled by the average 
cerebral capacity. The importance of cerebral capacity in our model suggests that incorporation
of this notion into theoretical and empirical studies of cognitive processes can potentially be 
very beneficial.

The model studied here aims to describe only some aspects of the early stages of the 
evolution of intelligence. The model should not be applied directly to actual human history and 
society. Our model does not aim to explain why a cognitive explosion has occured only in the 
lineage leading to modern {\em Homo sapiens}. Rather it tests whether a particular
set of explanations advanced and discussed in detail by many  \citep{hum76,byr88,ale90,whi97,dun98,dun03,fli05,str05,gea05,rot05}, 
which places special emphasis on the achievement of ``ecological dominance'' and on 
competition in regard to social competencies, is
plausible from the population genetics perspective. Alternative explanations do exist
(e.g., \citealt{vrb95,bin99,rus04}), and much more work remains necessary to better understand
the origins of human uniqueness.
As with most other mathematical models used in evolutionary biology
(e.g., \citealt{fis30,hal32,wri69,kim83,ewe79,bur00,gav04}), the goal of our model is not to
{\em prove} that a particular phemomena arises as a result of particular factors. Rather we wished
to explore the logic and plausibility of the arguments used to explain the phenomenon, 
to identify important factors, parameters, and time-scales, and to check the robustness of conclusions 
to variation in assumptions.

Our model has a number of limitations. Here we discuss some potential consequences of their
violation. We concentrated on a single population of small size. Allowing for more populations
connected by migration should accelerate the onset of cognitive explosion by increasing
the amount of new genetic variation. Once the cognitive explosion is initiated in a local
population, emigrants (males or females) will quickly spread their genes across the whole
system. We allowed only for positive Machiavellian fitness  $\mu$ of memes. If memes with
both positive and negative values of $\mu$ are possible, the process of cognitive explosion 
is expected to be delayed as deleterious memes will occasionally spread through the population
like an epidemic reducing the fitness advantage of having high cognitive abilities.  
We assumed that memes are copied with no regard to the fitness or status of individuals they are learnt
from. Selective imitation when memes are more likely to be learned from high fitness/status 
individuals, should accelerate the evolution of brain size. This 
expectation is supported by the fact that a behavior analogous to cognitive explosion was observed 
in a much simpler and less realistic model of selective imitation \citep{hig00}  
formalizing an integrant of Blackmore's ``big brain'' hypothesis \citep{bla99}. We did not allow for errors in 
meme copying. If such errors can only decrease the Machiavellian fitness of memes,
then the process of cognitive explosion will be slowed down. However if copying errors resulting
in meme improvement are possible, we expect higher Machiavellian fitnesses to be 
achieved which potentially can accelerate the process. 
We conclude that overall, from the theoretical perspective, 
the phenomenon of cognitive explosion, its patterns, and time-scales identified here appear to be robust.

Finally we note that the modeling framework we have developed can potentially be used to study
the evolution of languages \citep{now02} and the coevolution of genes and culture in 
general \citep{cav81,boy85}.

\section{Methods}

Here we provide some additional details on the model and simulations.

{\bf Constant viability selection.}\quad
Viability (i.e., the probability to survive to the age of reproduction) of a child with trait 
values $a$ and $c$ is
	\[
		v =\exp 
		\left\{ -0.5 \left[\left( \frac{a}{\sigma_a}\right)^2
		+\left( \frac{c/c_{\max}}{\sigma_c}\right)^2\right]\right\},
	\]
where $\sigma_a$ and $\sigma_c$ are parameters
measuring the strength of viability selection. 
The individual's death rate $d$ is a function of $v$ to be specified below.

{\bf Frequency-dependent selection for mating success in males.}\quad
The probability that a contest between males $i$ and $j$ with Machiavellian fitnesses
$m(i)$ and $m(i)$ is won by male $i$ is
	\[
		p(i,j)=\frac{ \exp\left\{ \gamma \left[m(i)-m(j)\right]\right\}}
		{1+\exp \left\{\gamma \left[m(i)-m(j)\right]\right\}},
	\]
where $\gamma$ is a scaling parameter
measuring how effectively an advantage in $m$ is translated into larger value of $p$. 
With this parameterization $p(j,i)=1-p(i,j)$
and the effects of memes known to both contestants cancel out. The expected proportion
of contests won by male $i$ is $p_e(i) = \sum_{j,j \neq i} p(i,j)/(N_m-1)$,
where $N_m$ is the number of males in the population. The male's mating rate is 
a function of $p_e(i)$ to be specified below.

{\bf Events.}\quad 
There are five types of events: birth and death of individuals and invention,
loss, and replication of memes. We say that an event occurs at rate $x$ if the 
probability of this event during a short time interval $dt$ is $xdt$.

Each female gives birth at a constant rate $b$. 
Male $i$ is chosen to be the father with a probability proportional to his mating
group size
	\[
	f(i)=f_{\min}+(f_{\max}-f_{\min})\ p_e(i)^{\lambda},
	\]
where $f_{\min},f_{\max}$, and $\lambda$ are parameters. Note that
if $p_e(i)=1$ (i.e., the male wins all contests), $f(i)=f_{\max}$, and
if $p_e(i)=0$ (i.e., the male looses all contests), $f(i)=f_{\min}$.
If one defines a parameter $f_0$ as the mating group size at $p_e(i)=1/2$ 
(which is the case when everybody has the same Machiavellian fitness), then
	\[
	\lambda= \ln \left( \frac{f_{\max}-f_{\min}}{f_0-f_{\min}} \right)/\ln 2.
	\]
Parameters $f_{\min}, f_0$, and $f_{\max}$ can be thought of as the effective number of
females available to a male in the corresponding category. One can set $f_0=1$ with
$f_{\min}<1<f_{\max}$.
If a birth is to take place, a single offspring is produced
with account of recombination, segregation, and mutation. The sex is assigned randomly.
Then viability selection follows and surviving offspring instantaneously become adults.


Adults die at rate $d=N/K$, where $N$ is the overall population size and $K$ is the
population carrying capacity.

Males invent new memes at a constant rate $\nu$.
The values of Machiavellian fitness $\mu$ and complexity $\pi$ to be assigned to a new meme are
drawn randomly from a truncated bivariate normal distribution with constant 
means $\overline{\mu}=0.5, \overline{\pi}=0.5$, standard
deviations $\sigma_{\mu}$ and $\sigma_{\pi}$, and positive correlation $\rho$. 
Only the values satisfying the conditions $0<\mu<1,\pi_{\min}<\pi<1$, where $\pi_{\min}$ 
is a minimum meme complexity, are allowed.

Each meme is forgotten at a constant rate $\delta$.

%
Consider a meme with complexity $\pi$ present in the population in $M$ copies.
Consider also a male with learning ability $a$ and cerebral capacity $c$ who has already learnt $n$
other memes. The rate at which the male aquires the new meme is 
$\eta \frac{ a}{\pi} \exp \left[ - \beta \left( \frac{n}{c} \right)^{\gamma} \right] M$,
where 
$\eta$, $\beta$, and $\gamma$ are positive scaling parameters. The exponential term describes the
brain's saturation with memes. Note that if $\gamma$ is large, then this term is either
close to 1 (if $n<c$) or close to zero (if $n>c$).



{\bf Simulations.}\quad The model dynamics are simulated using Gillespie's direct method \citep{gil77}. 
That is, the next event to happen is chosen according to the corresponding rates. The time 
interval until the next event is drawn from an exponential distribution with a parameter 
equal to the sum of the rates of all possible events. All rates are recomputed after each event.  

{\bf Initial conditions and parameters.}\quad
Initially, all individuals are identical homozygotes with $a=c=0$
and no memes.
We varied the number of loci per trait ($L=8,16,32$),
the population carrying capacity ($K=50, 100,150$),
the maximum cerebral capacity ($c_{\max}=16, 32, 64$),
the maximum mating group size ($f_{\max}=5, 10, 20$), and the
correlation between meme Machiavellian fitness and complexity ($\rho=0.25, 0.5, 0.75$).
The following parameters did not change:
mutation probability  per locus $10^{-5}$,
$\sigma_a=\sigma_c=2$, $\gamma=0.5,\beta=1, \gamma=10$, $b=2.2, f_{\min}=0$, 
$\nu=0.01, \delta=0.02, \eta=0.05$, $\sigma_{\mu}=\sigma_{\pi}=0.25$, $\pi_{\min}=0.05$.
40 runs were done for each of 243 parameter combinations.
Simulations ran for 30,000 time units (roughly corresponding to 30,000 generations)


\vspace{1cm}
{\small ACKNOWLEDGMENTS. 
We thank J. H. Williams, H. N. Qirko,  and A. Kramer  for discussions and comments,
B. M. Fitzpatrick for help with statistical analysis, and M. D. Vose for help with
numerical algorithms. 
Supported by the National Institutes of Health and by 
the National Science Foundation. The simulations were done on the 
Frodo and Grig clusters of the Scalable Intracampus Research Grid
(SInRG), University of Tennessee, Knoxville.}\newline

\bibliography{kniga,mach}

\end{document}